\documentclass[letterpaper,twocolumn,english,pra,aps]{revtex4}
\usepackage[T1]{fontenc}
\usepackage[latin1]{inputenc}
\usepackage{graphicx}

\makeatletter

\@ifundefined{definecolor}
 {\usepackage{color}}{}

\makeatletter

\makeatletter

\usepackage{soul}

\makeatother

\makeatother

\makeatother

\usepackage{babel}

\begin{document}

\title{Ghost imaging with a single detector}

\author{Yaron Bromberg$^{\dagger}$}

\affiliation{Department of Physics of Complex Systems, The Weizmann
Institute of Science, Rehovot, Israel}

\author{Ori Katz$^{\dagger}$}

\affiliation{Department of Physics of Complex Systems, The Weizmann
Institute of Science, Rehovot, Israel}

\author{Yaron Silberberg}

\email[]{yaron.silberberg@weizmann.ac.il}

\affiliation{Department of Physics of Complex Systems, The Weizmann
Institute of Science, Rehovot, Israel}
\begin{abstract}
We experimentally demonstrate pseudothermal ghost imaging and ghost
diffraction using only a single single-pixel detector. We achieve
this by replacing the high resolution detector of the reference beam
with a computation of the propagating field, following a recent
proposal by Shapiro [J. H. Shapiro, arXiv:0807.2614 (2008)]. Since
only a single detector is used, this provides an experimental
evidence that pseudothermal ghost imaging does not rely on non-local
quantum correlations. In addition, we show the depth-resolving
capability of this ghost imaging technique.
\end{abstract}


\maketitle \noindent Ghost imaging (GI) has emerged a decade ago as
an imaging technique which exploits the quantum nature of light.
This field has attracted great interest, and has been in the focus
of many studies since. In GI an object is imaged event hough the
light which illuminates it is collected by a single-pixel detector
which has no spatial resolution (a bucket detector). This is done by
using two spatially correlated beams. One of the beams illuminates
the object, and the photons transmitted by the object are collected
by the bucket detector. The other beam impinges on a multipixel
detector (e.g. a CCD camera), without ever passing through the
object (the reference beam). Nevertheless, by correlating the
intensities measured by the bucket detector with the intensities of
each pixel in the multipixel detector, an image of the object is
reconstructed \cite{GattiReview}. In a similar fashion, the
diffraction pattern of the object can also be obtained ('ghost
diffraction', GD). In the first demonstrations of GI and GD the two
beams were formed from a stream of entangled photons
\cite{PitmanPRA95,FirstGD95}. The reconstruction of the image was
attributed to the non-local quantum correlations between the photon
pairs. Challenging this interpretation, Bennink \textit{et al.}
demonstrated GI using two classically correlated beams
\cite{Boyd02}, and triggered an ongoing effort to clarify the role
of entanglement in GI and GD
\cite{GattiPRL03,GattiPRL04,Boyd04,GattiPRL05,ShihPRL05,ShihPRL06,
Comment2Shih,OttonelloOEX07,ShapiroPRA08_GauusianGI,ShihSoldierPRA08}.
It was soon discovered that many of the features obtained with
entangled photons, are reproduced with a classical pseudothermal
light source \cite{GattiPRL04,GattiPRL05,ShihPRL05}. However, the
nature of the spatial correlations exhibited with a pseudothermal
source, and whether they can be interpreted as classical intensity
correlations \cite{Comment2Shih,GattiReview,ShapiroPRA08_GauusianGI}
or are fundamentally non-local quantum correlations
\cite{ShihPRL06,ShihReview}, is still under debate.

In this Letter we experimentally study \textit{computational ghost
imaging}, a novel ghost imaging technique recently proposed by
Shapiro \cite{shapiroCGI}. In this technique the multipixel detector
is replaced with a 'virtual detector', by calculating the
propagation of the field of the reference beam. The image is
reconstructed by correlating the calculated field patterns with the
measured intensities at the object arm. Our measurements show that
pseudothermal GI and GD can be performed with only one beam and one
detector. As noted by Shapiro, this proves that GI cannot possibly
depend on any non-local quantum correlations. We also demonstrate
scanningless 3D sectioning capability of this technique.

\begin{figure}
\centering{}\includegraphics[width=1\columnwidth]{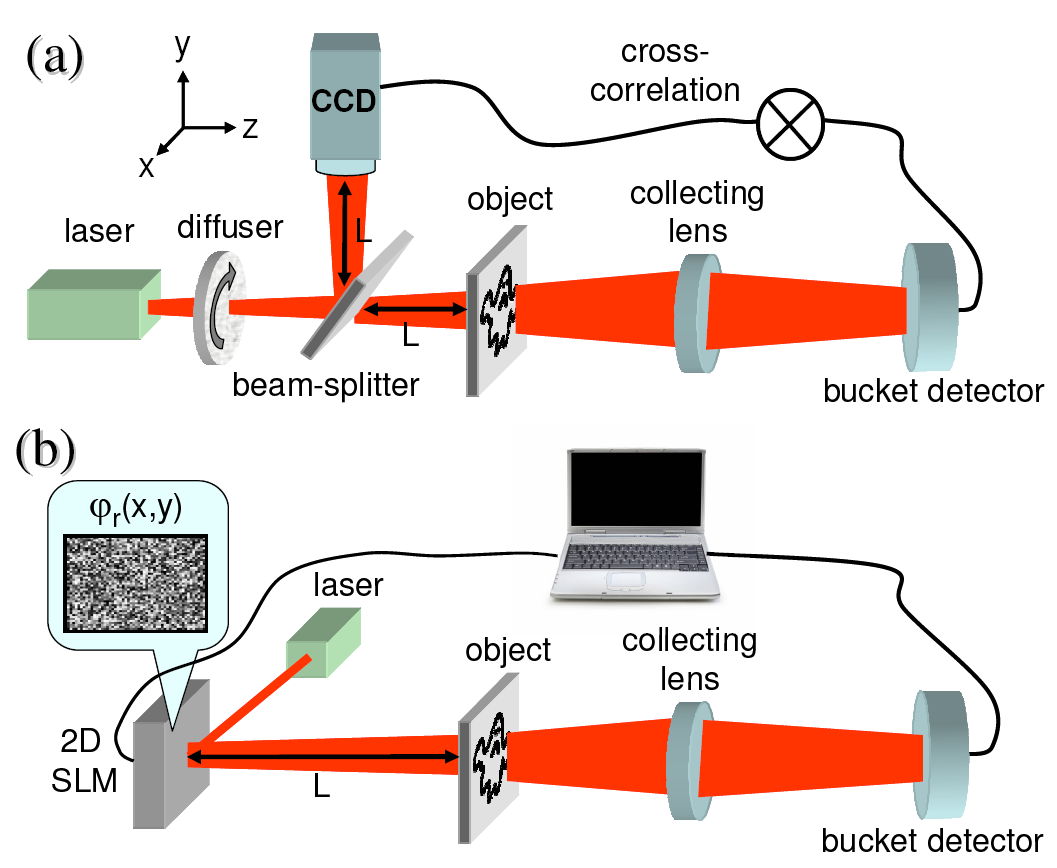}

\caption{\label{fig:setup} (Color online) Experimental setups for
ghost imaging: (a) The standard pseudothermal two-detectors setup,
where a ghost image of the object is obtained by correlating the
pseudothermal field measured by a CCD with the intensity measured by
a bucket detector. (b) The computational single-detector setup used
in this work. A pseudothermal light beam is generated by applying
controllable phase masks $\varphi_{r}(x,y)$ with a spatial light
modulator (SLM). The object image is obtained by correlating the
intensity measured by the bucket detector, with the
\emph{calculated} field at the object plane.}

\end{figure}

Computational GI can be considered as a variant of the standard
two-detectors pseudothermal GI. In pseudothermal GI, a spatially
incoherent beam is generated by passing a laser beam through a
rotating diffuser (Fig. \ref{fig:setup}a). The beam is then split on
a beam-splitter, generating the two spatially correlated beams
required for GI. The essence of computational GI is to replace the
rotating diffuser with a computer controlled spatial light modulator
(SLM), which serves as a controlled phase-mask for the spatial phase
of the light field $\varphi(x,y)$ (Fig.~\ref{fig:setup}b). A
spatially incoherent beam is generated by applying pseudo-random
phase patterns $\varphi_{r}(x,y)$ on the SLM. Since for each phase
realization $r$ the controlled phase pattern is known, one can
evaluate the field right after the SLM,
$E_{r}(x,y,z=0)=E^{(in)}e^{i\varphi_{r}(x,y)}$ (where $E^{(in)}$ is
the incident field on the SLM). Knowing $E_{r}(x,y,z=0)$, the field
at any distance $z$ from the SLM can be computed using the
Fresnel-Huygens propagator:

\begin{equation}
E_{r}(x,y,z)=\int d\xi d\eta
E_{r}(x-\xi,y-\eta,0)e^{i\frac{\pi}{\lambda z}
(\xi^{2}+\eta^{2})},\label{eq:fresnel}
\end{equation}

\noindent where $\lambda$ is the wavelength of the source. In order
to reconstruct the transmission function of an object $T(x,y)$ placed
at $z=L$, the \textit{computed} intensity patterns at the object
plane $I_{r}=|E_{r}(x,y,z=L)|^{2}$ are cross-correlated with the
intensities \textit{measured} by the bucket detector placed behind
the object $B_{r}=\int dxdyI_{r}(x,y,L)T(x,y)$:

\begin{eqnarray}
G(x,y) & \equiv & \frac{1}{N}\sum_{r=1}^{N}(B_{r}-\langle B\rangle)I_{r}(x,y)\label{eq:Gxy}\\
 & = & \langle BI(x,y)\rangle-\langle B\rangle\langle I(x,y)\rangle.\nonumber \end{eqnarray}

\noindent Where $\langle\cdot\rangle\equiv\frac{1}{N}\sum_{r}\cdot$
denotes an ensemble average over $N$ phase realizations. Intuitively,
one can see that the image is obtained by summing the calculated intensities
$I_{r}$ with the appropriate weights $B_{r}$: The larger the overlap
between the generated intensity pattern and the transmission object,
the higher is the intensity measured by the bucket detector $B_{r}$,
and thus the calculated $I_{r}(x,y)$ is summed with a larger weight.
It is important to note that in conventional GI the object is also
reconstructed according to Eq.~(\ref{eq:Gxy}). The only difference
is that here $I_{r}(x,y,z=L)$ are \textit{computed}, whereas in conventional
GI $I_{r}$ are obtained by measuring the intensities at the reference
arm using a multipixel detector placed at a specific distance $z=L$.
Thus, similar to what is well known for conventional GI, one can show
that $G(x,y)$ is given by the transmission function of the object
convolved with the coherence function of the field at the plane of
the object \cite{GattiReview,ShapiroPRA08_GauusianGI}. The resolution
of the reconstructed object is therefore limited by the coherence
area of the field at the object plane.

\begin{figure}
\centering{}\includegraphics[width=1\columnwidth]{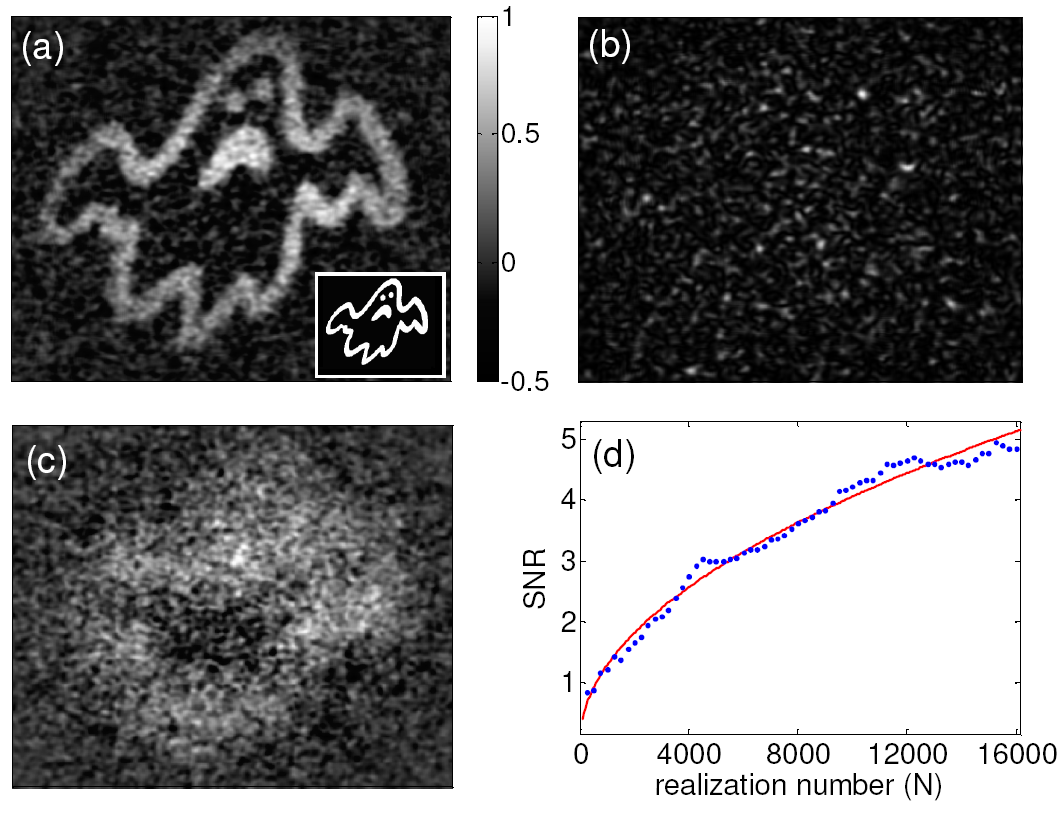}

\caption{\label{fig:image}(Color online) Computational ghost image
reconstruction of a $2cm\times2cm$ transmission mask placed at
$L=84cm$. (a) Reconstructed image at the object plane, obtained with
$16000$ realizations. The inset shows the transmission mask. (b) A
calculated intensity pattern of a single phase realization. The
resolution of the reconstruction in (a) is dictated by the speckle
size. (c) Reconstructed out-of-focus image, at a different z-plane
($L=15cm$), demonstrating the depth-resolving capabilities of the
computational method. (d) Measured SNR of the reconstructed image as
a function of the number of realizations (blue dots). The
theoretical line depicts $\sqrt{N}$ dependence. }

\end{figure}

To demonstrate computational GI experimentally, we have constructed
the setup presented in Fig. \ref{fig:setup}b. The setup is based
on a two-dimensional phase-only liquid crystal on silicon spatial-light-modulator
(LCOS-SLM, Holoeye HEO1080P), with 1080x1920 addressable $8\mu m\times8\mu m$
pixels. The computer controlled 2D-SLM is illuminated by a CW Helium-Neon
laser, which produces a Gaussian beam with a waist of $w_{0}=740\mu m$
on the 2D-SLM plane. A $2cm\times2cm$ object (transmission plate)
is placed at distance $L=84cm$ from the SLM, and a lens collects
the transmitted light onto the bucket detector. In each realization
a mask of $300\times300$ random phases is sent to the SLM, where
each phase is realized by 3x3 SLM pixels. The reconstructed image
using $16000$ realizations is shown in Fig. \ref{fig:image}a, displaying
the accurate reconstruction of the object transmission $T(x,y)$ (Fig.
\ref{fig:image}a, inset). This experimental result cannot be attributed
to nonlocal quantum correlations, since only a single detector was
used. We note that one can reconstruct an object placed at any distance
$L$ from the source, both in the near- and far-field zones, as long
as the field at the object plane can be calculated using Eq. (\ref{eq:fresnel}).
This is demonstrated below, with an object placed at $L=11cm$.

An intensity pattern calculated from a \textit{single} realization
is shown in Fig. \ref{fig:image}b, revealing the speckle field that
impinges on the object for this specific realization of random phases
\cite{Goodman}. Reconstruction of an image at a z-plane different
than the actual location of the object, results in an out-of-focus
image of the object, indicating the depth-resolving capabilities of
computational GI (Fig. \ref{fig:image}c). This is analogous to scanning
the z-position of the multipixel detector in conventional GI, but
is performed here for all z-locations simultaneously, without the
need for additional measurements. This feature allows for a full three-dimensional
reconstruction of the object field, with the only price of executing
more calculations.

Since the computational GI scheme is completely analogous to pseudothermal
GI, the two techniques share similar resolution and signal-to-noise
(SNR) properties \cite{Goodman,GattiPRA08,ShapiroSNR,shapiroCGI}.
To summarize these, the transverse resolution is determined by the
coherence length (speckle size) at the object plane, which is given
by the Van-Cittert Zernike theorem $\delta x(z)=\lambda z/\pi w_{0}$.
The depth resolution is given by $\delta z(z)=2\pi\delta x^{2}/\lambda$.
For the presented experimental parameters, $\delta x(L=84cm)=230\mu m$
and $\delta z(L=84cm)=50cm$, in agreement with the experimental results.
In pseudothermal GI, the SNR scales as the square-root of the ratio
of the number of realizations $N$ and the average number of speckles
transmitted through the object $N_{s}$, $SNR\propto\sqrt{N/N_{s}}$
\cite{ShapiroSNR,GattiPRA08}. Since $N_{s}$ is given by the ratio
of the object transmissive area to the coherence area, there is a
clear trade-off between resolution and SNR. The SNR as a function
of the number of realizations for the reconstructed image of Fig.
\ref{fig:image}a, is presented in Fig. \ref{fig:image}d, in agreement
with the theoretical prediction. A movie visualizing the image build-up
can found at \cite{movie}.

\begin{figure}
\centering{}\includegraphics[width=1\columnwidth]{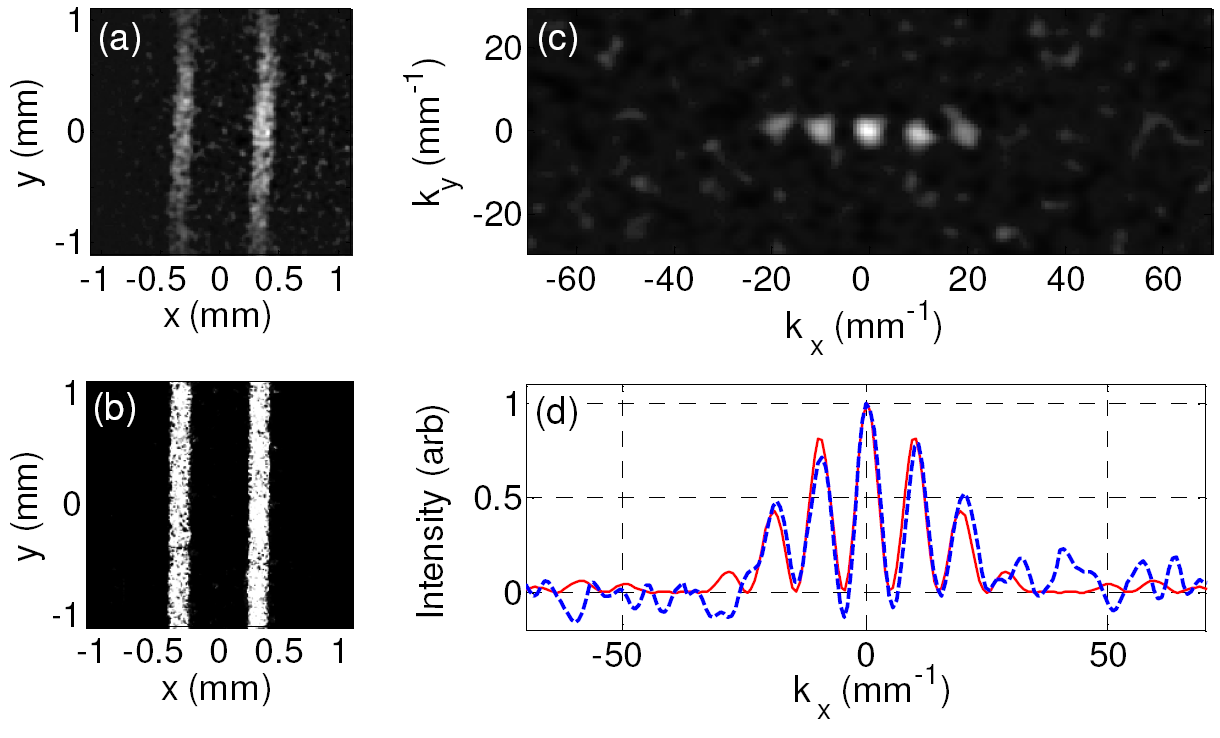}

\caption{\label{fig:diffraction} (Color online) Simultaneous reconstruction
of an object and its diffraction pattern, with $8000$ realizations.
(a) Reconstructed ghost image of a two-slit transmission plate. The
resolution of the image is determined by the speckle size, $\delta x\sim0.018mm$
(b) Transmission microscope image of the double-slit slide. The width
of each slit is $170\mu m$ and the separation is $400\mu m$. (c)
Ghost diffraction pattern reconstructed using a pinhole detector.
The resolution of the diffraction image is $\delta k\sim1.4mm^{-1}$,
lower than the Fourier limit $\delta x\delta k<0.5$. (d) A cross-section
of the ghost diffraction pattern shown in (c) (dashed blue), and the
theoretical two-slit diffraction pattern calculated from the dimensions
of the double-slit (red). }

\end{figure}

One of the features in common to ghost imaging with a nonclassical
source and a pseudothermal source, is the ability to resolve both
the object and its diffraction pattern with high resolution. In conventional
GD, the diffraction pattern of the object is reconstructed by replacing
the bucket detector with a small single-pixel ('pinhole') detector,
placed at the Fourier plane of the collecting lens \cite{Boyd04,GattiPRL05}.
The GD is reconstructed by correlating the intensities measured by
the pinhole detector with the diffraction pattern of the field in
the reference arm. In fact, both the near-field object image (GI)
and its diffraction pattern (GD) can be obtained by changing only
the optical setup at the reference arm. In our single detector configuration,
since the reference arm is virtual, the only required change is in
the computational procedure. One can therefore perform GI and GD simultaneously,
with a single set of measured data (i.e. the intensities measured
with the pinhole detector $B_{r}$). Both images are reconstructed
using Eq. (\ref{eq:Gxy}), where for GI $I_{r}$ is computed according
to Eq. (\ref{eq:fresnel}), and for GD $I_{r}$ is obtained by calculating
the Fourier transform (FT) of $E_{r}(x,y,z=0)$, $I_{r}=|FT(E_{r}(x,y,z=0))|^{2}$.
The intensities measured by a pinhole detector placed on the optical
axis are given by $B_{r}=|\int dxdyE_{r}(x,y,L)T(x,y)|^{2}$. We note
that by using a pinhole detector instead of a bucket detector, the
SNR of the GI is degraded, since only a small fraction of the transmitted
light is collected.

In order to demonstrate the simultaneous reconstruction of GI and
GD, we have placed a double-slit transparency at $L=11cm$ from the
SLM and a small pinhole detector at the Fourier plane of the
collecting lens (a single $8.6\mu m\times8.4\mu m$ pixel of a CCD
camera). We reconstructed the double-slit diffraction pattern using
the pinhole-detector measured intensities, and its transmission
image using a bucket-detector measured intensities (summing over the
CCD's pixels), obtained simultaneously with the same phase
realizations. We have also used the pinhole detector to reconstruct
both the GD and GI, but with much lower SNR for the latter (not
shown). Fig.~\ref{fig:diffraction} summarizes the results of the
near-field and far-field image reconstructions for the double slit
transparency, obtained with $8000$ realizations. Comparing
Fig.~\ref{fig:diffraction}a and Fig.~\ref{fig:diffraction}d, it is
obvious that the product of the resolution of the near-field image
$(\delta x)$ and of the far-field image $(\delta k\sim1/w_{0})$ is
much smaller than the Fourier limit, i.e. $\delta x\delta
k=0.025\ll0.5$. This cannot be obtained by simply scanning the
object with a coherent laser beam \cite{Boyd04}. However, it does
not violate the Einstein-Podolsky-Rosen (EPR) bound for classical
light, as was previously shown with a pseudothermal source
\cite{GattiPRL05}. Knowing both the near- and far-field intensity
patterns with high resolution enables the reconstruction of the
object's transmission amplitudes \emph{and} phases, using the
Gerchberg Saxton phase-retrieval algorithm \cite{Saxton72}. This
ability of computational GI might be appealing for phase sensitive
applications.

In conclusion, we have shown that pseudothermal GI can be performed
with only a single detector, thus proving that it does not rely on
non-local quantum correlations. In addition, we have demonstrated
the depth resolving and diffraction imaging capabilities of this
single-detector technique. Finally, we note that in computational GI
the main complexity is shifted from the experimental apparatus to
the calculation. Thus, it allows for 3D reconstruction by only post
processing the retrieved data, eliminating mechanical scans. One
might consider applying computational GI for other 3D imaging tasks
which are not necessarily transmission-based. Two examples for such
are radar applications, and scanningless depth-resolved microscopy
using fluorescent probes. Furthermore, the use of a SLM enables the
implementation of closed-loop feedback schemes, potentially reducing
the number of realizations needed for image reconstruction.

\end{document}